\documentclass[11pt,a4paper]{article}

\usepackage{jheppub}

\renewcommand{\d}{\textrm{d}}
\newcommand{\e}{\textrm{e}}

\title{Non-SUSY fractional branes}

\author[a]{Stanislav Kuperstein,} \author[b]{Brecht Truijen} \author[b]{and Thomas Van Riet}

\affiliation[a]{Institut de Physique Th\'eorique, CEA
 Saclay, CNRS URA 2306,  F-91191 Gif-sur-Yvette, France} 

\affiliation[b]{Instituut voor Theoretische Fysica, K.U. Leuven, Celestijnenlaan 200D B-3001 Leuven, Belgium} 

\emailAdd{stanislav.kuperstein@gmail.com}\emailAdd{brecht.truijen@fys.kuleuven.be}\emailAdd{thomasvr@itf.fys.kuleuven.be}

\abstract{We consider a simplified Ansatz for supergravity solutions describing fractional $p$-brane solutions (throat geometries supported by fluxes) for various $p$. For $p=3$ the Ansatz captures the Klebanov-Tseytlin (KT) solution. The equations of motion can be derived from an effective action by performing a dimensional reduction to a flat domain wall geometry in $p+2$ dimensions.  We find an interesting  deformation of the known superpotential defining the SUSY domain wall flow.  The deformation parameter breaks supersymmetry but still preserves the property that a test D$p$ brane feels no force inside the throat. The new solutions come in two classes. Both classes have the same UV asymptotics as the KT solution and one class has also similar IR behavior, which makes them potentially interesting holographic backgrounds for studying cascading gauge theories with broken SUSY.  We explain furthermore  how the curved domain wall solutions of the same ($p+2$)-dimensional theories are expected to lift to new AdS compactifications of type IIA/B supergravity.}

\preprint{IPhT-T14/181}

\begin{document}
\maketitle

\setcounter{footnote}{0}
\setcounter{figure}{0}
\setcounter{equation}{0}

\section{Introduction}

One of the pressing problems in string phenomenology and formal string theory is to increase our understanding of controlled mechanisms to break supersymmetry (SUSY). In the supergravity limit of string theory one can attempt to construct explicit solutions describing SUSY-breaking phenomena. A popular method relies on adding SUSY-breaking branes (anti-branes) to SUSY backgrounds whose warping can red-shift the SUSY-breaking sources and hence control their back-reaction \cite{Maldacena:2001pb, Kachru:2002gs, Argurio:2006ny, Kachru:2003aw}. 
In the context of the AdS/CFT correspondence such supergravity solutions might capture the elusive dynamical SUSY-breaking in strongly coupled gauge theories.  Away from the decoupling limit, in compact models, the previously proposed methods for SUSY-breaking could lead to a landscape of meta-stable de Sitter vacua in string theory \cite{Kachru:2003aw}.

Recently these SUSY-breaking mechanisms have been the subject of heavy scrutiny because of certain, potentially unphysical, singularities found in the supergravity solutions, see for instance \cite{Bena:2009xk, McGuirk:2009xx,Bena:2012bk} for the earliest comments. These singularities arise because of a no-force condition that is being violated: the fluxes in the throat geometries feel a force that draws them towards the SUSY-breaking brane at the tip of the warped throat \cite{Blaback:2010sj, Blaback:2011nz, DeWolfe:2004qx}. Indeed, the specific form of the singularity is such that the density in the fluxes diverges near the source. The orientation of the fluxes is furthermore such that it describes an infinite pile-up of brane charge dissolved in fluxes with a sign opposite to that of the SUSY-breaking brane.  The force on the fluxes can be thought of as the force on a test D-brane inside the throat since fluxes act in a way very similar to a smooth distribution of branes.

One reason to worry about the singular flux density is that high flux densities, when they grow indefinitely, trigger perturbative brane-flux annihilation, turning the background perturbatively unstable \cite{Blaback:2012nf, Danielsson:2014yga}. To circumvent these issues we are interested in solutions that break SUSY but at the same time still preserve the no-force condition on the fluxes or test D$p$-branes. This is by no means a unique way to circumvent the singularities inherent to anti-brane solutions, but the no-force condition is one of the reasons why supersymmetric backgrounds are easier controllable. Non-SUSY solutions sharing the no-force property might be the first place to look for controlled SUSY-breaking. 

In this paper we are able to find  \emph{first-order} (``fake SUSY like") equations for such solutions. It is intriguing that we can find explicit first-order flows that break SUSY but preserve the no-force condition. As it turns out, our new solutions indeed evade the singular flux-pile up that are typical to solutions that break the no-force condition. The case of $p=3$ is analysed in some detail and the SUSY solution is the so-called Klebanov-Tseytlin (KT) solution \cite{Klebanov:2000nc}. The solutions we find are distortions of the KT solution that are controlled by a deformation parameter. The UV behavior of the solutions is identical to KT and some of the new solutions have also a very similar IR region, which contains a naked singularity. In case SUSY is not broken it is well known that this singularity can be resolved either by deforming the conifold \cite{Klebanov:2000hb} or by putting the system at finite temperature \cite{Aharony:2007vg}. We leave the question whether the singularity can be resolved for future work, though we initiate an attempt to describe the finite temperature solution through first-order equations. 

Non-SUSY no-force solutions have already occurred in the literature. Such a background was for instance constructed in \cite{Dymarsky:2011ve}, but the approach of \cite{Dymarsky:2011ve} differs from ours since it leads to second-order equations of motion.

This paper is organised as follows. In Section \ref{sec:Technique} we review the relation between first-order flow equations, Hamilton-Jacobi theory and fake SUSY applied to domain wall flows in theories of gravity coupled to scalars subject to some scalar potential. We use this formalism to describe the new SUSY-breaking fractional branes in Section \ref{sec:branes}. To apply our formalism to this set-up we first dimensionally reduce the 10-dimensional geometry to a flat domain wall Ansatz. We discuss the absence of a force on a test D$p$-brane inside the throat. Section \ref{finite} extends the analysis to finite temperature and we conclude in Section \ref{sec:concl}. In Appendix A we briefly discuss the curved (AdS-sliced) domain walls of the theories we have investigated and their lift to ten dimensions.

\section{Effective actions and first-order flows}
\label{sec:Technique}

In gravity theories, coupled ordinary first-order  equations (flow equations) can arise whenever the
Ans\"atze contain sufficient spacetime symmetries (e.g. spherical black holes, FLRW cosmologies, domain walls). The equations of motion can then be recast into ordinary differential equations of the standard Hamiltonian form which can be derived from a one-dimensional effective action:
\begin{equation} 
\label{Lagrangian}
 I = \int \d\rho \Bigl( \frac{1}{2} G_{ij}(q) \dot{q}^i\dot{q}^j - \mathcal{V}(q)\Bigr)\,.
\end{equation}
The canonical variables $q^i$ ($i=1,\ldots,N$) are the undetermined functions of a single coordinate, $\rho$, in the solution Ans\"atze. The conditions for preserving some amount of SUSY then become first-order flow equations, typically of the form of an autonomous system:
\begin{equation}\label{firstorder}
\dot{q}^i =  F^i(q_1,\ldots, q_N)\,,
\end{equation}
where $F^i$ are certain functions of the canonical variables. Such first-order integrations can sometimes be found for solutions that break SUSY as has been observed over the last 10 years for black holes, domain walls, cosmologies and general, non-SUSY, $p$-brane solutions. In fact, many of these examples are related by dimensional reduction, which then relate the corresponding first-order equations. When a $p$-brane is reduced over its spatial worldvolume it becomes a black hole ($0$-brane). If, instead, it is reduced over the angles in the transversal space it becomes a domain wall.  

It has been observed by several authors that the first-order equations (\ref{firstorder}) are instances of the Hamilton-Jacobi equation in classical mechanics \cite{deBoer:1999xf,  Skenderis:2006rr, Andrianopoli:2009je, Trigiante:2012eb}. This implies that the first-order equations (\ref{firstorder}) are equivalent to gradient flow equations:
\begin{equation}
\label{HJ0}
\dot{q}^i = G^{ij} \frac{\partial \mathcal{S}}{\partial q^j}\,,
\end{equation}
where the gradient function $\mathcal{S}(q)$ is \emph{Hamilton's principal function}. In the supergravity literature this gradient function is sometimes called \emph{fake superpotential}. This is, however, a slight abuse of language since SUSY-like equations (\ref{firstorder}) are only equivalent to Hamilton-Jacobi equations when the principal function factorises \cite{Trigiante:2012eb} as further reviewed in Section \ref{sec:FSusy}. To avoid confusion, from here on we will use the abbreviation $\mathcal{S}$-function.

\subsection{Algebraic Hamilton-Jacobi theory} \label{Section:Legendre}

We first revise some aspects of Hamilton-Jacobi (HJ) theory.
Thanks to diffeomorphism covariance it is always possible to parametrise the gravitational Ans\"atze of interest such that neither the metric $G_{ij}$ nor the potential $\mathcal{V}$ in (\ref{Lagrangian}) have explicit $\rho$-dependence. Hamilton's principal function then takes the form $\mathcal{S}(q^i)-E \rho$, where the first term is $\rho$-independent and $E$ is a constant, which has the interpretation of energy for ordinary dynamical systems.\footnote{For a separated solution $\mathcal{S}(q^i,t)=\mathcal{S}^\prime(q^i)-Et$, the function $\mathcal{S}^\prime$ is usually called the \emph{characteristic} function. In all of the example we will consider, however, $E=0$, and so we will still refer both to $\mathcal{S}$ and $\mathcal{S}^\prime$ as the principal function and use the same notation for the two of them.} The  $\mathcal{S}$-function then satisfies the following Hamilton-Jacobi (HJ) equation:
 \begin{equation}
 \label{eq:HJ}
\frac{1}{2}G^{ij} \frac{\partial \mathcal{S}}{\partial q^i}  \frac{\partial \mathcal{S}}{\partial q^j} + \mathcal{V} = E\,.
 \end{equation}

The non-dynamical part of the Einstein equations enforces an algebraic ``zero energy" constraint for the gravitational systems of interest. For FLRW cosmologies and domain walls this nicely translates into $E=0$ in (\ref{eq:HJ}). Spherical black holes, on the other hand, have $E=0$ for extremal solutions and non-zero $E$ for non-extremal ones, where $E$ becomes proportional to $ST$, the product of entropy and temperature.\footnote{To be more precise, $E$ is zero both for extremal and non-extremal solutions but the non-extremal black holes have a constant term in the potential of (\ref{eq:HJ}) which can then be reinterpreted as the energy \cite{Janssen:2007rc}.}
 
The canonical momenta $p_i$ associated to the canonical variables in (\ref{Lagrangian}) are defined by $p_i = G_{ij}\dot{q}^j$. The HJ equations of motion then take the form:
\begin{equation}
p_i = \frac{\partial \mathcal{S}}{\partial q^i} \, , 
\end{equation}
which are thus equivalent to the flow equations (\ref{HJ0}).

HJ theory implies that a first-order integration always exists locally. These flow equations (\ref{HJ0}) may be integrated to obtain a solution to the equations of motion. However, the original purpose of HJ theory is not to integrate first-order equations but rather solve them algebraically. Recall that the most general solution to the HJ equation depends on $N$ constants of motion $\alpha_i$. For convenience we identify $\alpha_N$ as $E$. The HJ formalism then implies that the quantities:
\begin{equation}
\label{eq:betas}
\beta^i \equiv \frac{\partial \mathcal{S} \left( q^k, \alpha_k \right)}{\partial \alpha_i}
\qquad i=1, \ldots, N \, ,
\end{equation} 
are also constants of motion, $\dot{\beta}^i=0$. Consequently (\ref{eq:betas}) and  (\ref{HJ0}) become a set of algebraic equations for $q^i(\rho)$ and $p_i (\rho)$ (or, equivalently $\dot{q}^i(\rho)$) in terms of $\alpha_i$'s and $\beta^i$'s, which are in turn determined by the initial conditions. The physical meaning of the constants $\alpha_i$'s and $\beta^i$'s is easy to understand by recalling that Hamilton's principal function is in fact a generating function for a certain canonical transformation $\left( q^i, p_i\right) \to \left( Q^i, P_i\right)$. The HJ equation (\ref{eq:HJ}) for $\mathcal{S}$ then implies that the new Hamiltonian $H^\prime \left(Q^i, P_i\right)$ identically vanishes, guaranteeing that $\dot{P}=0$ and $\dot{Q}=0$. If the $\mathcal{S}$-function depends on $q^i$'s and $P_i$'s, then $\alpha_i$ and $\beta^i$ correspond to $P_i$ and $Q^i$ respectively.


In practice, finding a complete solution of the HJ equation (\ref{eq:HJ}) is rather rare for gravity systems and one can typically only hope to find the dependence on a few $\alpha_i$'s out of the full set of $N$ constants. If, for instance, $\mathcal{S} (q^i, \alpha_1)$ is available, then we end up with one algebraic equation, namely (\ref{eq:betas}) for $i=1$, together with $N-1$ independent first-order differential equations (\ref{HJ0}), which we still have to solve.

For completeness, let us mention that starting from $\mathcal{S} \left( q^i, \alpha_i \right)$ we can always introduce a new generating function $\widetilde{\mathcal{S}} \left( q^i, \beta^i \right)$ by performing the Legendre transformation on the $\alpha_i$'s:
\begin{equation}
\label{Legendre}
\widetilde{\mathcal{S}} \left( q^i, \beta^i \right) = \mathcal{S}  \left( q^i, \alpha_i \right) - \beta^j \alpha_j \, ,
\end{equation}
where $\beta^i$ on the right hand side is defined by (\ref{eq:betas}).\footnote{One can verify that the new $\mathcal{S}$-function solves the original HJ equation simply by noticing that
\begin{equation}
\left. \frac{\partial \mathcal{S}}{\partial q^i} \right\vert_{\alpha_j} = \left. \frac{\partial \widetilde{\mathcal{S}}}{\partial q^i} \right\vert_{\beta_j} \, ,
\end{equation}
which also shows that the physical meanings of $\alpha_i$'s and $\beta^i$'s is reversed under the transformation.} Although the new principal function, $\widetilde{\mathcal{S}} \left( q^i, \beta^i \right)$, seems to provide a new solution of the HJ equation, the final solution of the equations of motion is merely a reparametrization of the (most general) solution obtained using the original principal function $\mathcal{S}  \left( q^i, \alpha_i \right)$.

\subsection{Relation with fake supersymmetry } \label{sec:FSusy}

The prime goal of this subsection is to clarify the connection between the superpotential approach and the Hamilton-Jacobi formalism applied to gravity systems. The discussion is based on \cite{deBoer:1999xf, Trigiante:2012eb}. 

Consider the following action of gravity coupled to $n$ scalar fields in $D$ spacetime dimensions:
\begin{equation}\label{actionscalars}
S = \int\sqrt{-g}\Bigl\{R - \frac{1}{2}h_{ab}\partial\phi^a\partial\phi^b - V(\phi) \Bigr\}
\qquad
a,b=1,\ldots,n
\, .
\end{equation}  
Such theories can support non-trivial domain wall geometries which are solutions to the following metric Ansatz:
\begin{equation} \label{domainwall}
\d s^2_D =  f(z)^2\d z^2  + g(z)^2 \d\Sigma_k^2\,,
\end{equation} 
where the $D-1$ dimensional metric $\Sigma_k^2$ is either Minkowskian ($k=0$), AdS ($k=-1$) or dS ($k=+1$). In what follows we consider Minkowski-sliced domain walls, also known as flat domain walls. The scalar fields are considered to depend on the coordinate $z$ which parametrizes the distance from the wall located at some fixed value $z=z_0$. The function $f(z)$ in the metric is a gauge choice and can (locally) be chosen at will. It is well known that for any function $\mathcal{W}$ that obeys
\begin{equation} \label{fs1}
V = \frac{1}{2} h^{ab}\partial_a \mathcal{W} \partial_b \mathcal{W}  - \frac{D-1}{4(D-2)}\mathcal{W}^2\,, 
\end{equation} 
the following first-order flows \cite{DeWolfe:1999cp} (where a dot denotes a $z$-derivative):
\begin{eqnarray}
 f^{-1}\dot{\phi}^a &=& - h^{ab}\partial_b \mathcal{W}\,,\label{fs2} \\
 f^{-1}\frac{\dot{g}}{g} &=& \frac{\mathcal{W}}{2(D-2)} \,, \label{fs3}
\end{eqnarray}
solve the second-order equations of motion. This function $\mathcal{W}$ is called a \emph{fake superpotential} when it is not related to supersymmetry and for supersymmetric solutions it corresponds to the genuine \emph{superpotential}.
 
As discussed in various papers \cite{deBoer:1999xf, Andrianopoli:2009je, Trigiante:2012eb, Skenderis:2006rr} these first-order equations are a manifestation of the Hamilton-Jacobi formalism for one-dimensional Hamiltonian systems. The Hamiltonian system is defined by the second-order differential equations, which can be derived from the following effective action obtained from the Ans\"atze (\ref{actionscalars}), (\ref{domainwall}):
\begin{equation}
\label{effaction}
I = \int \d \rho \, \left\{ \frac{g^{D-1}}{f}\left( -\frac{1}{2}h_{ab}\dot{\phi}^a\dot{\phi}^b + (D-1)(D-2)\left( \frac{\dot{g}}{g} \right)^2 \right)  - f g^{D-1} V \right\} \,.
\end{equation} 
The variable $f$ is algebraic (since it stems from a degree of freedom that can be gauged away) and it enforces the zero energy constraint $E=0$. One must choose a gauge for $f$ to regard this as a Hamiltonian system in $n+1$ variables $g, \phi^i$ and impose the zero energy constraint separately. When Hamilton's principal function $\mathcal{S}(q)$ factorises as follows \cite{Trigiante:2012eb}:
\begin{equation}
\label{factorisation1}
\mathcal{S} = g^{D-1}\mathcal{W}(\phi) \, ,
\end{equation} 
the HJ equations (\ref{HJ0}, \ref{eq:HJ}) with $E=0$ then reproduce exactly the fake SUSY first-order flow equations (\ref{fs1}-\ref{fs3}) independently of the chosen gauge for $f$.\footnote{Notice that $f$ in (\ref{effaction}) can be absorbed in a proper $\rho$-coordinate redefinition.}  When such solutions have a holographic dual, the first-order HJ equation (\ref{HJ0}) is dual to the RG-flow equations of the  gauge theory \cite{deBoer:1999xf}
Whether this factorisation can always occur is a subtle issue \cite{Trigiante:2012eb} and is still under investigation \cite{future}.

\section{Fractional branes as domain walls}\label{sec:branes}

In this section we study Ans\"atze that captures $p$-branes magnetically charged under the $F_{8-p}$ RR field strength along with extra NSNS $H_3$ flux and RR $F_{6-p}$ flux. The latter fluxes can be combined into a $(9-p)$-form $H_3\wedge F_{6-p}$ that acts as a smooth magnetic source together with the singular $p$-brane source:
\begin{equation}
\d F_{8-p} = H_3\wedge F_{6-p} + \delta_{9-p} (\textrm{D}p)\,.
\end{equation}
Such solutions play a prominent role for non-conformal holographic backgrounds or, when the D$p$-brane is replaced by an O$p$-plane \cite{Blaback:2012mu}, serve as standard flux compactifications. 
Maybe the most well known example is the Klebanov--Strassler geometry \cite{Klebanov:2000hb} for $p=3$, but here we study more symmetric examples for $p=3$ \cite{Klebanov:2000nc} and for general $p$ \cite{Herzog:2000rz, Janssen:1999sa, Blaback:2012mu}.

The way we proceed is to first reduce the brane solutions over the angles in their transversal space to a flat domain wall in $p+2$ dimensions. This allows us to write down the effective action (\ref{Lagrangian}) in a simple manner and in doing so we also make contact with the concept of fake SUSY, which was historically developed for the special case of domain walls \cite{Freedman:2003ax}. We find new first-order flow equations that preserve one interesting property of the SUSY solutions: the force on a probe D$p$ brane inside the throat vanishes for any position of the probe. 

Not every new domain wall flow that solves the equations of motion is necessarily a physical solution. For domain walls in 5D (i.e. $p=3$) we investigate the solutions in some details and observe that they all posses a naked singularity at small radius whereas the large radius behaviour is identical to the SUSY Klebanov-Tseytlin (KT) solution. The KT solution also has a naked singularity but it is known to be resolved when the conifold is deformed \cite{Klebanov:2000hb}, yielding the Klebanov-Strassler (KS) solution. There exists a simple criterion \cite{Gubser:2000nd} for naked singularities to be admissible. Assuming that the singularities have certain brane interpretation, at finite temperature the solution must develop a horizon. Hence, the criterion states that admissible singularities must be cloaked by a horizon at sufficiently high temperature. This will briefly be discussed in Section \ref{finite}.

In the next subsection we present the 10D Ansatz of IIA/IIB SUGRA that describes our solutions.

\subsection{Dissolved fractional branes (branes in throats)}
In 10D Einstein frame we consider the following Ansatz:
\begin{equation}
\mathrm{d}s^2_{10}=e^{2A(\rho)}\mathrm{d}s^2_{p+1}+e^{2B(\rho)} \mathrm{d}s^2_{9-p}\,.
\label{eq:DpAnsatz}
\end{equation} 
The worldvolume metric is assumed to be flat, $\d s_{p+1}^2 =\eta_{\mu\nu}\d x^{\mu}\d x^{\nu}$, whereas the transversal space is assumed to be a cone over a base space, $\Sigma_{8-p}$, with the following metric:
\begin{equation}\label{metric0}
\mathrm{d}s^2_{9-p}=\mathrm{d}\rho^2+e^{2C(\rho)}\left[ g_{ij}^{\Sigma}(\psi)\mathrm{d}\psi^i\mathrm{d}\psi^j\right],\\
\end{equation} 
where $\rho$ is the radial coordinate and the metric $g^\Sigma_{ij}$ is assumed to be Einstein:
\begin{equation}
R^{\Sigma}_{ij} = (7-p)g^{\Sigma}_{ij} \, ,
\end{equation} 
such that for $e^{2C}=\rho^2$ the metric $\d s^2_{9-p}$ is Ricci flat. The field content that supports the metric is given by:
\begin{eqnarray}
\label{eq:DpAnsatz2}
H_3 &=& \d B_2 = b'(\rho) \mathrm{d}\rho \wedge \epsilon_2 \nonumber\\
F_{6-p} &=& m \star_{8-p} \epsilon_2 \\
F_{8-p} &=& Q \epsilon_{8-p} + B_2 \wedge F_{6-p} \,,
\nonumber
\end{eqnarray}
where $\star_{8-p}$ is the Hodge star on the base space $\Sigma_{8-p}$,  $m$ is some constant which measures the fractional fluxes and $\epsilon_2$ is a 2-form on the cone base.  The constant $Q$ can be absorbed in the definition of $b(\rho)$ but it will be useful to keep $Q$ explicit in order to discuss the limit $m\rightarrow 0$ where it becomes related to the brane charge. The field equations require that the two-form $\epsilon_2$ is closed and co-closed. The fields only depend on $\rho$. 
As usual there is the subtlety that the Ansatz for $F_5$ has to be changed to:
\begin{equation}\label{F5}
F_5 = (1+\star_{10})(Q \epsilon_{5}  + B_2 \wedge F_{3} )\,.
\end{equation} 
 
The known solutions are given by \cite{Herzog:2000rz, Blaback:2012mu}:
\begin{eqnarray}
\d s^2_{10} & = & H^{\frac{p-7}{8}} \d s^2_{p+1}(x^a) +
H^{\frac{p+1}{8}}\d s^2_{9-p}(y^i)\,,\label{metric}\\
\e^{\phi} & = & g_s H^{\frac{3-p}{4}}\,, \label{phi}\\
F_{8-p} & = & -g_s^{\frac{3-p}{4}} \star_{9-p}(\partial_i H \d
y^i)\,, \label{F_{8-p}}\\
F_{6-p} & = & g_s^{-\frac{p+1}{4}}\star_{9-p} H_3 \label{F_{6-p}}\,,
\end{eqnarray}
where, again, the transversal and world-volume metrics are Ricci flat. The Hodge star $\star_{9-p}$ is defined with respect to the flat metric $\d s^2_{9-p}(y^i)$.
For $p= 1,2,4,6$, we have explicitly:
\begin{equation}\label{KH1}
H =  C_2 + \frac{\tilde Q}{r^{7-p}} -
\frac{g_s^{\frac{p-1}{2}}\,m^2}{2(p-3)(p-5)}\,\frac 1 {r^{10-2p}}\,,
\end{equation}
where $C_2$ is arbitrary (taken zero in the decoupling limit). $\tilde Q$ determines the charge of the $p$-brane at the tip (but such an interpretation is subtle) and, up to factors and constants, it agrees with the $Q$ introduced in (\ref{eq:DpAnsatz2}). We use $r$ as the notation for the radial coordinate, instead of $\rho$, when we employ the gauge choice 
\begin{equation}
e^{2C(r)}=r^2 \,,
\label{eq:RadialR}
\end{equation} 
in the cone-metric (\ref{metric0}).
For asymptotically (Ricci) flat backgrounds we can set $C_2=1$. 

According to \cite{Herzog:2000rz} there exist no solutions with $p=5$. For $p=3$
the last term becomes a logarithm:
\begin{equation}\label{KH2}
H =C_2 + \frac{\tilde Q}{r^4} + \frac{g_s m^2}{16\, r^4}\Bigl( 4 \,\ln r +
1\Bigr)\,.
\end{equation}
Supersymmetry requires the choice of a suitable
conical internal space. For D3 branes ($p=3$) the metric $\d\Omega_5$ should be that of the conifold $T^{1,1}$ and this solution was first described by Klebanov and Tseytlin in \cite{Klebanov:2000nc} and will be referred to as KT in what follows.

For $p=6$ no SUSY is  possible with this Ansatz. A more general solution for $p=6$ was found in \cite{Janssen:1999sa} and is characterised by the following
choice for $H$:
\begin{equation}\label{JMO1}
H(r, z) = 1 + \frac{\tilde Q}{r} -\frac{1}{6}g_s^{\frac{5}{2}} m^2 r^2 + c z\,,
\end{equation}
where $z$ is a Cartesian spatial coordinate on the D6 worldvolume,
$m$ is the Romans mass and $c$ a constant. When $m$
and $c$ both vanish this is the standard extremal D6 solution in IIA
supergravity, which preserves 1/2 of the supersymmetry. When $m\neq
0$ it was shown that maximally 1/4 of the supersymmetry could be
preserved if  $c=m$. 

\subsection{The lower-dimensional theory}
The metric on the ($9-p$)-dimensional space transversal to the brane can be written in polar coordinates and for $p$-branes with sufficient symmetries the fields only depend on the radial coordinate and not on the angles. The angles themselves parameterise a space $\Sigma_{8-p}$ with Einstein metric  of positive curvature. It is natural to perform a dimensional reduction over the Einstein space such that the $p$-brane has but a single transversal direction (the radial direction) in the reduced theory that lives in $p+2$ spacetime dimensions. This way $p$-branes reduce to domain walls. If the reduced theory is a maximal gauged supergravity, the 1/2 SUSY domain walls lift to 1/2 SUSY $p$-branes in the $10$-dimensional theory, see \cite{Bergshoeff:2004nq} for a list of explicit examples. 

The reason for reducing the 10-dimensional brane solutions to flat domain walls in $p+2$ dimensions is mainly practical; the Kaluza--Klein reduction offers a useful form for the effective action for the unfixed degrees of freedom in the action. It also allows us to make contact with the concept of fake supersymmetry, which is usually defined for domain wall space-times. Although in this paper we will only barely touch upon the issue of fake supersymmetry.

The Ansatz for a dimensional reduction over $\Sigma_{8-p}$ is simply a rewriting of the metric Ansatz (\ref{eq:DpAnsatz})  for the fractional brane solutions:
\begin{equation}\label{simpleAnsatz}
\d s^2_{10} = e^{2\alpha \varphi} \d s^2_{p+2} + e^{2\beta\varphi}\d\Sigma_{8-p}^2\,.
\end{equation}
This is not the most general Ansatz but describes a consistent truncation for three dynamical scalar fields in $p+2$ dimensions: $\phi, \varphi, b$. If we choose 
\begin{equation}
\beta = -\frac{p\alpha}{8-p}\,,\qquad \alpha^2 = \frac{8-p}{16p}\,,
\end{equation} 
the reduced theory is in Einstein frame and the scalar $\varphi$ is canonically normalised. The reduced action is:
\begin{equation} \label{action}
I  = \int \sqrt{-g_{p+2}}\Bigl\{ R_{p+2} -\frac{1}{2}(\partial\varphi)^2 -\frac{1}{2}(\partial\phi)^2 - \frac{1}{2}e^{\frac{4p\alpha}{8-p}\varphi -\phi}(\partial b )^2 - V(\varphi,\phi, b)\Bigr\} \,,
\end{equation} 
with the following scalar potential:
\begin{eqnarray}
V(\varphi,\phi, b)  &=&  -(8-p)(7-p)e^{\frac{16\alpha}{8-p}\varphi} + \frac{1}{2}m^2 e^{\frac{2\alpha}{8-p}(8+5p-p^2)\varphi +\frac{(p-1)}{2}\phi} \nonumber\\ 
&& + \frac{1}{2}(Q+mb)^2 e^{2(p+1)\alpha\varphi  + \frac{(p-3)}{2}\phi}\,.\label{potential}
\end{eqnarray}

\subsubsection*{Flat domain walls}

If we take a flat domain wall metric (\ref{domainwall}) for $\d s_{p+2}^2$ one can indeed fit the fractional D$p$ solutions when uplifted to $10d$. The known SUSY fractional D$p$ solutions, we discussed in (\ref{metric} - \ref{F_{6-p}}) can be obtained from the following principal function:
\begin{equation} \label{S0}
\mathcal{S}_0(g, \phi, \varphi, b) = g^{p+1}\Bigl((Q+mb)\,e^{\frac{(p-3)}{4}\phi +(p+1)\alpha\varphi} \pm 2(8-p) e^{\frac{8}{8-p} \alpha\varphi}\Bigr)\,.  
\end{equation} 
This principal function is indeed of the fake SUSY form (\ref{factorisation1}) such that the term between brackets defines the superpotential $\mathcal{W}(\phi,\varphi, b)$ (\ref{factorisation1}):
\begin{equation}
\mathcal{W}(\phi,\varphi, b) = (Q+mb)\,e^{\frac{(p-3)}{4}\phi +(p+1)\alpha\varphi} \pm 2(8-p) e^{\frac{8}{8-p} \alpha\varphi}\,.
\end{equation} 
One can explicitly check that $\mathcal{W}$ obeys (\ref{fs1}) with $D=p+2$ for the system defined in (\ref{action}), (\ref{potential}).

\subsubsection*{Curved domain walls}

One could also wonder about the curved (AdS-sliced) domain walls of the reduced theories in $p+2$ dimensions. Since we have not yet found first-order flow equations for all $p$ (for $p=6$ this has already been found in \cite{Apruzzi:2013yva}). We rather discuss the would-be solutions in general in the appendix to demonstrate how their lift predicts a generalisation of a set of warped \emph{compact} AdS vacua of recent interest \cite{Blaback:2011nz, Blaback:2011pn, Apruzzi:2013yva, Junghans:2014wda, Gaiotto:2014lca, Danielsson:2013qfa}.

\subsection{Non-SUSY solutions} 
\label{sec:NewSol}
The main result of this paper is that we have found a second \emph{family} of solutions of the Hamilton-Jacobi equation:
\begin{equation} 
\label{S1}
\mathcal{S}_1(c) = g^{p+1} \left(\, (Q+mb) \, \,e^{\frac{(p-3)}{4}\phi +(p+1)\alpha\varphi}  \pm 2(8-p) \, e^{\frac{8}{8-p} \alpha\varphi} \cosh \left( \gamma_1 \phi +\gamma_2 \varphi + \gamma_3 \ln(g)  + c \right) \right)\,,
\end{equation}
where: 
\begin{equation}
\gamma_1=\frac{(7-p)}{4}\sqrt{\frac{p+1}{2(8-p)}}\, ,\qquad \gamma_2 = -4\frac{(p-3)}{(7-p)}\gamma_1\alpha  \,,\qquad \gamma_3=\frac{\gamma_2}{\alpha} \,,
\end{equation} 
and $c$ is an arbitrary real constant. Obviously $c$ is an example of the constant $\alpha_1$ we introduced in Section \ref{sec:Technique}.

It is not immediately clear that the principal function $\mathcal{S}_1(c)$ describes solutions that can be seen as one-parameter deformations of the SUSY solutions described by $\mathcal{S}_0$ since there is no limit for the integration constant $c$ for which $\mathcal{S}_1(c)$ coincides with $\mathcal{S}_0$. However, we can look at the Legendre transform, (\ref{Legendre}), which generates an equivalent $\mathcal{S}$-function, $\mathcal{S}_2(d)$, as explained in Section \ref{Section:Legendre}:
\begin{eqnarray} \label{S2}
&& \mathcal{S}_2(d) = (Q+mb) \,g^{p+1} \,e^{\frac{(p-3)}{4}\phi +(p+1)\alpha\varphi} \pm 2\sqrt{d^2+(8-p)^2 g^{2(p+1)}e^{\frac{16}{8-p}\alpha \varphi}} \\
 && \,  +2d\left[ \gamma_1 \phi + (\gamma_2 + \frac{8\alpha}{8-p})\varphi + (\gamma_3 +p+1)\ln(g) \right] - 2d\ln\left(  \sqrt{d^2+(8-p)^2 g^{2(p+1)}e^{\frac{16}{8-p}\alpha \varphi}} \pm d\right) \, ,
 \nonumber
\end{eqnarray}
where $d$ is analogous to $\beta_1$ in (\ref{eq:betas}). 
This one-parameter family $\mathcal{S}_2(d)$, with the continuous parameter $d$, is connected to the SUSY solutions described by $\mathcal{S}_0$ by taking $d=0$, i.e. \mbox{$\mathcal{S}_2(d=0)= \mathcal{S}_0$}. The first-order flows from $\mathcal{S}_1(c)$ and $\mathcal{S}_2(d)$ can be mapped to each other by substituting $c$ as a function of $d$ and the other fields: $c=c(d, \phi, \varphi, b, g)$. 

In the absence of 3-form fluxes ($m=0$) we can obtain explicit solutions and they coincide with the non-SUSY brane solutions constructed earlier by Lu and Pope in \cite{Lu:1996hh}. For the latter solutions it was indeed observed that they could be found from a first-order order formulation \cite{Janssen:2007rc} and our results can be seen as an extension that includes extra form fluxes. In the decoupling limit of the $p=3$ solution we expect that our new solutions coincide with the ones of \cite{Gubser:1999pk}.

\subsection*{(Fake) supersymmetry}
The known SUSY solutions are found from the principle function $\mathcal{S}_0$,\footnote{To be more precise, the solutions defined by $\mathcal{S}_0$ are SUSY for a proper choice of the metric $\d\Omega_{8-p}$. } hence the new principle functions $\mathcal{S}_1(c)$ (or equivalently $\mathcal{S}_2(d)$) give rise to non-SUSY solutions. We therefore can verify whether the new solutions are fake SUSY, prior to solving the first-order Hamilton-Jacobi flow equations. Earlier we mentioned that fake SUSY is usually defined as a factorisation of the principle function (\ref{factorisation1}).
The principle function $\mathcal{S}_1(c)$ in (\ref{S1}) factorises in the sense (\ref{factorisation1}) when $p=3$ such that we may naively speak of a fake superpotential. However, the fully equivalent principal function $\mathcal{S}_2$ in (\ref{S2}) does not factorise and one would correspondingly not speak of fake SUSY. This confusion is related to the subtlety pointed out in \cite{Trigiante:2012eb} and can be resolved by a more careful definition of the concept of fake supersymmetry, which is the subject of future work \cite{future}.

\subsection*{Explicit analysis for $p=3$}

We have not been able to find analytic solutions for all fields to the first-order flow equations for these new principal functions for generic values of $c$. However, having the first-order equations at hand a numerical integration becomes rather straightforward. We now present some analytic and numerical results for $p=3$ from the $10d$ point of view for easy comparison to the known fractional brane solutions \cite{Klebanov:2000nc, Herzog:2000rz, Blaback:2012mu}. We focus on solutions in the decoupling limit, by which we mean that the large $r$ behavior corresponds with that of solution (\ref{KH2}) with $C_2=0$.

To compare the new solutions with the known SUSY ones (\ref{KH2}), we are interested in the behaviour of $A$, $B$, $b$ in  (\ref{eq:DpAnsatz}, \ref{eq:DpAnsatz2}) and the dilaton $\phi$. We employ the standard gauge for the radial coordinate $r$ in (\ref{eq:RadialR}), used in (\ref{KH2}).\footnote{This corresponds to the choice $f(r)=r^{-1}e^{\frac{8\beta}{p}\varphi}$.} The flow equations derived from (\ref{S0}, \ref{S1}, \ref{S2}) can be rewritten to the $10d$ field basis, for $\mathcal{S}_0$ this becomes:
\begin{eqnarray}
\label{eq:S0flow}
A^\prime &=& \frac{(Q+mb)e^{-4B}}{4r^{5}}\qquad , \qquad b^\prime = m\frac{e^{\phi}}{r}  ,
\\
B^\prime &=&-\frac{(Q+mb)e^{-4B}}{4r^{5}} \quad \, , \qquad \phi^\prime=0. 
\end{eqnarray}
Here a prime indicates differentiation w.r.t.~$r$. This is the SUSY case and it is clear that we can impose Ricci flat conditions at large $r$. The flow can be integrated to give (\ref{metric}, \ref{phi}, \ref{F_{8-p}}, \ref{F_{6-p}}, \ref{KH2}). The non-SUSY flows, defined by $\mathcal{S}_1(c)$ only differ in the equations for $B$ and $\phi$ while the equations for $A$ and $b$ remain unchanged.  Using $\mathcal{S}_1(c)$ one finds:
\begin{eqnarray}
\label{eq:S1flow}
B^\prime &=& -\frac{(Q+mb)e^{-4B}}{4r^{5}} + \frac{1}{r} \left( \frac{1}{r} \cosh \left( \sqrt{\frac{2}{5}}\phi+c \right) - 1 \right) , \\
\phi^\prime &=&-\frac{2\sqrt{10} }{r} \sinh \left( \sqrt{\frac{2}{5}}\phi+c \right) \, .
\nonumber 
\end{eqnarray}
Additionally, HJ theory provides us with an algebraic equation (\ref{eq:betas}) relating the constants $c$ and $d$:
\begin{equation}
\label{eq:algebraicHJ}
d = -5 e^{4A + 4B}r^4 \sinh \left( \sqrt{\frac{2}{5}}\,\phi +c \right) \, .
\end{equation}
We will now solve the equations arising from $\mathcal{S}_1$ and compare it to the SUSY solutions (\ref{F5}, \ref{metric}, \ref{phi} \ref{KH2}). In our conventions we have $Q>0$ such that we end up with the standard D3 brane solution in the limit $m\rightarrow 0$.\footnote{A negative $Q$ could be interpreted as an orientifold solution since one obtains an orientifold solution in flat space in the limit $m=0$. } The non-SUSY dilaton equation can be straightforwardly integrated to give:
\begin{equation}
e^{\phi(r)}
= e^{\phi_\infty}  \left( \frac{r^4-r_s^4}{r^4+r_s^4} \right)^{\sqrt{\frac{5}{2}}}  \, .
\end{equation}
Here we identified $c$ in $\mathcal{S}_1$ with the value of the dilaton at $r \to \infty$: 
\begin{equation}
c=-\sqrt{\frac{2}{5}} \phi_\infty \, , 
\end{equation}
and $r_s$ is an integration constant corresponding the minimal value of the radial coordinate $r$. With this choice of the sign in front of $r_s^4$, the string coupling goes to zero in the IR, while for the opposite sign it diverges at $r=r_s$. In what follows we will stick with the former option. In the SUSY case the dilaton is obviously constant, which is reproduced for $r_s=0$.



Next, we solve for $e^{A+B}$ using the $A^\prime$ and $B^\prime$ equations in (\ref{eq:S0flow}) and (\ref{eq:S1flow}) respectively. For the SUSY background $A+B=0$ everywhere while the non-SUSY solution gives:
\begin{equation}
\label{eq:exp(A+B)}
e^{ 4 \left( A(r)+B(r) \right) } = 1- \left( \frac{r_s}{r} \right)^8 \, .
\end{equation}
This result suggests that we have a singularity at $r=r_s$ as long as $r_s$ does not vanish and we have verified that this is indeed the case. Again, for $r_s=0$ we reproduce the SUSY solution. One can verify that $d=10 r_s^4$.

The solution of the $b(r)$ equation of motion is simple but does not have a neat analytic form. For large $r$ the function $b(r)$ diverges as in the KT solution:
\begin{equation}
b(r) \approx m e^{\phi_\infty} \ln r + \ldots \, ,
\end{equation}
and near $r=r_s$ the function $b(r)$ is finite: 
\begin{equation}
b(r) = b_s + \mathcal{O} \left( \left( r - r_s \right)^{\sqrt{\frac{5}{2}}+1} \right)\,.
\end{equation}
Although $b_s$ is clearly a gauge constant and the full 3-form flux $H_3$ does not depend on it, a shift in $b_s$ nevertheless corresponds to a \emph{large} gauge transformation, since within our Ansatz \mbox{$\int_{S^2, r=\textrm{const}} B_2 = b(r) $}, where $S^2$ is the 2-sphere on the conifold base. The full $10d$ background will be therefore sensitive to the choice of $b_s$ similar to what happens in the SUSY KT geometry.

Finally we are in a position to study the behaviour of $e^{-4A}$ such that the full solution is understood. The relevant equation is:
\begin{equation}
\left( e^{- 4 A} \right)^\prime = - \frac{(Q+mb(r))}{r^5}  e^{-4(A+B)} =-(Q+mb(r))\frac{r^3}{r^8-r_s^8} \, .
\label{eq:4A}
\end{equation}
In the far UV ($r \to \infty$) the solution has the familiar form in agreement with the SUSY case:
\begin{equation}
 e^{- 4 A} \approx \frac{1}{4} m^2 e^{\phi_\infty} \cdot \frac{\ln r}{r^4} + \mathcal{O} \left( r^{-4}\right) \, ,
\end{equation}
but the IR behaviour ($r=r_s$) depends on the sign of $Q+mb(r_s)$. Loosely speaking the combination $Q+mb(r)$ can be interpreted as an effective charge density. Since the solutions are such that an effective no-force condition holds (see below), this effective charge is also an effective tension. Let us define $r_e$ as the radius where $Q+mb(r_e)=0$. At $r=r_e$ the effective tension of the ``would be" brane changes sign and gravity becomes repulsive. In the SUSY case the singularity is located at a radius smaller than $r_e$, let us call this radius $r_*$. The non-SUSY solutions, on the other hand, allow to shift the radius of the singularity $r_s$ such that both solutions with $r_e>r_s$ and $r_e<r_s$ exist. 

These two possibilities lead to a very different behaviour near the singularity. If we reach the singularity before $Q+mb(r)$ changes sign, $r_e<r_s$, then (\ref{eq:4A}) shows that $e^{-4A}$ diverges as:
\begin{equation}
e^{-4A} \approx -(Q+mb(r_s)) \ln (r-r_s) +\ldots \, .
\label{eq:Exp-4Adiverges}
\end{equation}
Note that while the metric is singular, the string coupling $e^\phi$, goes to zero at $r=r_s$. This behaviour is new and not present in the SUSY case.

If, on the contrary, $r_e>r_s$ then $Q+mb(r)$ turns negative in the IR and this forces $e^{-4A}\rightarrow 0$, even before we reach $r=r_s$. This behaviour is very similar to the SUSY case. The radius, $r_*(r_s, r_e)$, at which $e^{-4A}$ becomes zero is defined through: 

\begin{equation}
e^{-4A(r_*(r_s, r_e))} = \int_{r_*(r_s, r_e)}^{\infty} \mathrm{d}r (Q+mb(r))\frac{r^3}{r^8-r_s^8}=0 \,,
\label{eq:NewSingular}
\end{equation}
and satisfies $r_s<r_*(r_s,r_e)<r_e$. Note that (\ref{eq:NewSingular}) only makes sense if $Q+mb(r)$ becomes negative. This shows that whenever $r_e>r_s$, we find a singularity at $r_*(r_s, r_e)$ but, contrary to the $r_e < r_s$ case, the dilaton is well behaved (finite) at the singularity. This singularity is also very similar to the KT singularity of the SUSY case, which is confirmed by a numerical treatment which also shows that $r_*(r_s, r_e)$ agrees with $r_*$ in the limit $r_s\rightarrow 0$. The singularity at $r_s$ is no longer present as the geometry only exists for $r>r_*(r_s, r_e)$.

In the limiting case, $r_s=r_e$, the function $e^{-4A}$ approaches a finite value at $r=r_s$.

\begin{figure}[h!]
\includegraphics[width=\textwidth]{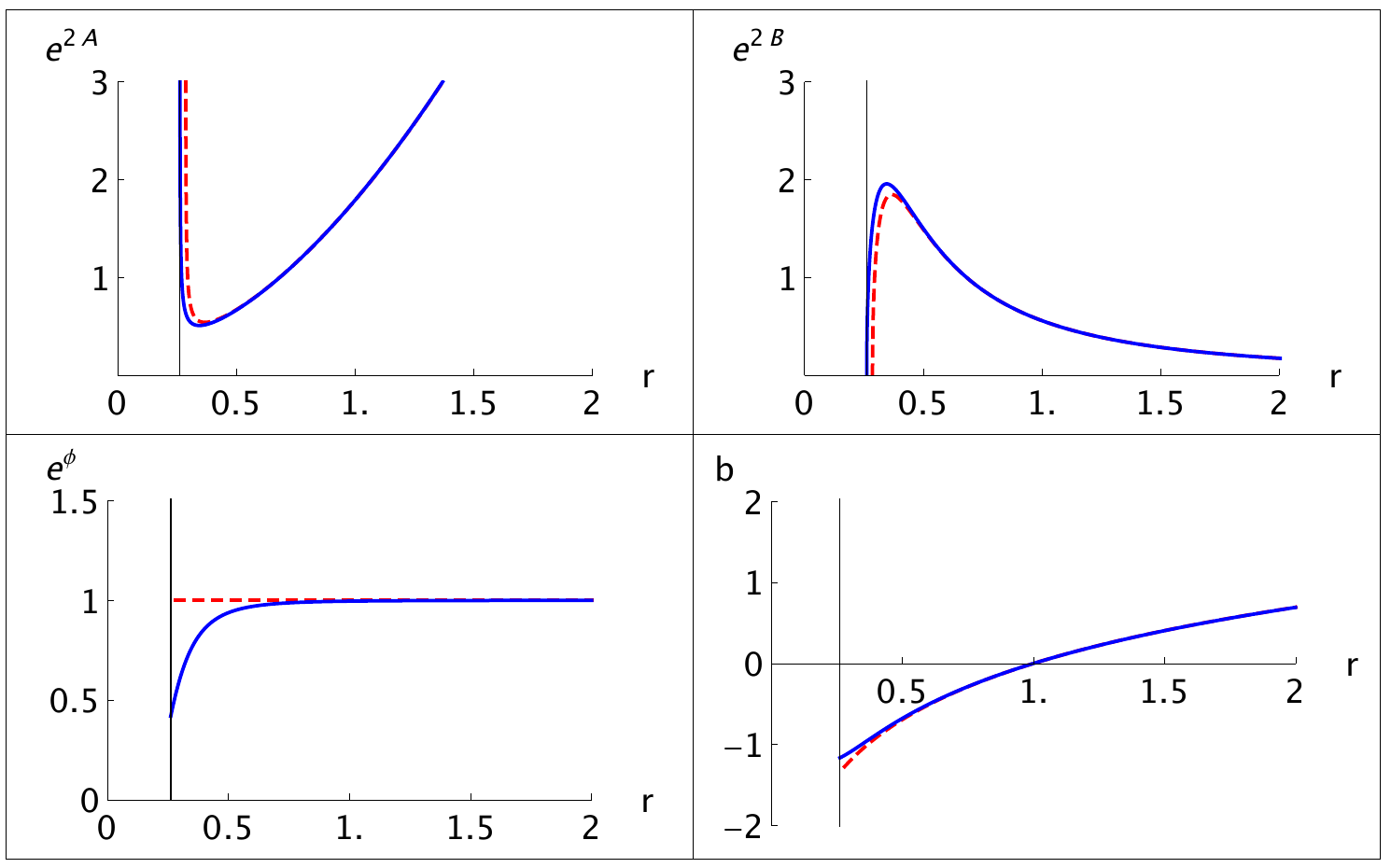}
\caption{\small \emph{Solutions to the flow equations which follow from $\mathcal{S}_0$ (dashed curve) and $\mathcal{S}_1$ (solid curve) with $r_e>r_s$ for $p=3$ and $m=1$, $Q=1$. 
Initial conditions where chosen such that the singularity of the SUSY case (dashed) is located at $r_*\approx 0,286505$ and $r_e\approx 0,367879$. For the non-SUSY solution (solid) we have $r_s\approx 0,188588$, $r_e\approx0,34466$ and a singularity forms at $r_*(r_s, r_e)\approx 0,262082$.  The vertical line denotes the point where the non-SUSY solution develops a singularity.}}
\label{fig:Plot1}
\end{figure}

\begin{figure}[h!]
\includegraphics[width=\textwidth]{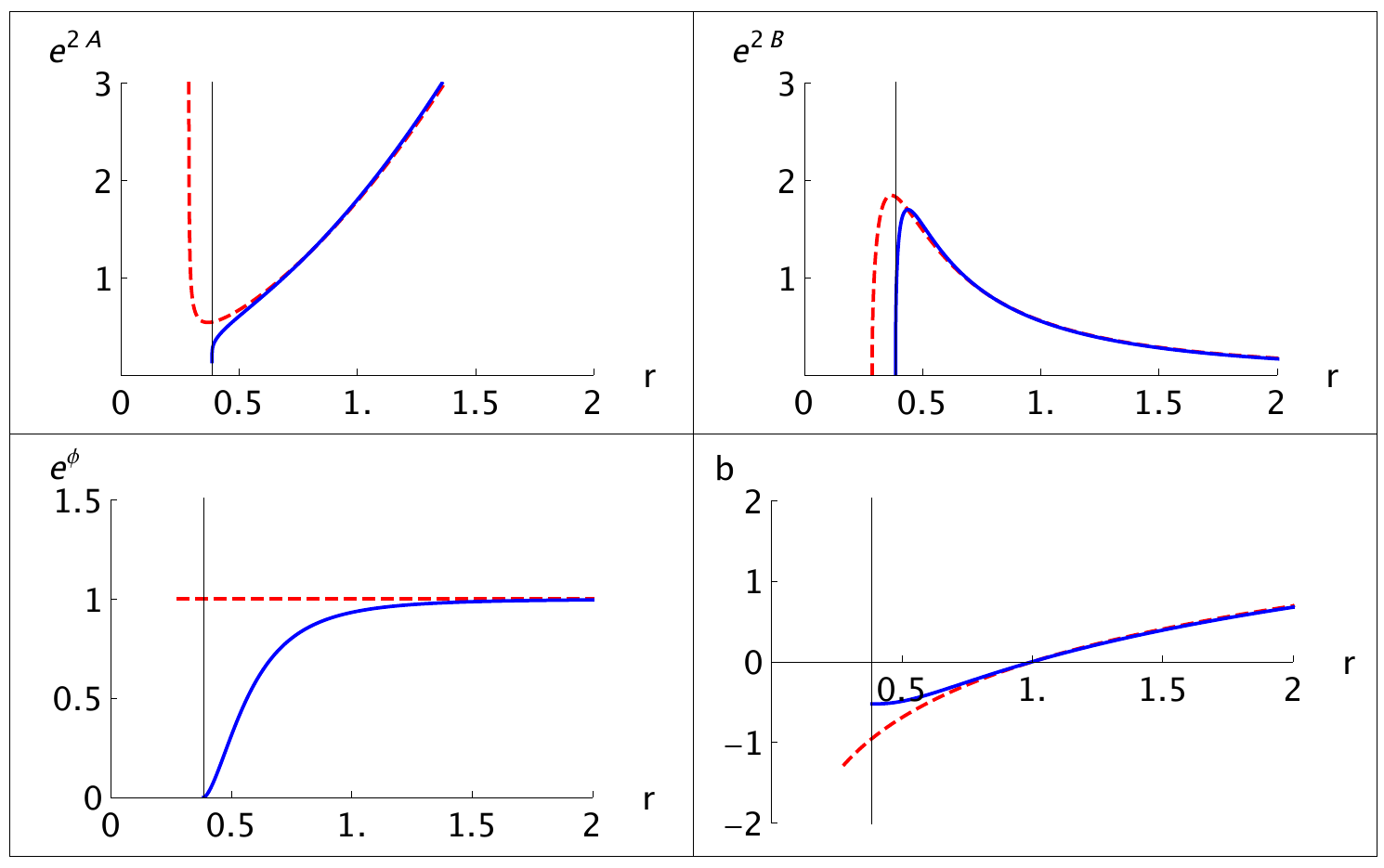}
\caption{\small \emph{The solutions to the flow equations which follow from $\mathcal{S}_0$ (dashed curve) and $\mathcal{S}_1$ (solid curve) with $r_e<r_s$ for $p=3$ and $m=1$, $Q=1$. 
Initial conditions where chosen such that the singularity of the SUSY case (dashed) is located at $r_*\approx0,286505$ and $r_e\approx0,367879$. For the non-SUSY solution (solid) we have $r_s\approx0,385706$ and $r_e\approx0,385133$.  The vertical line denotes the point where the non-SUSY solution develops a singularity.
}} 
\label{fig:Plot2}
\end{figure}

The above discussion is confirmed by a numerical treatment of the first-order flows. We focused on the solutions for which the string coupling decreases in the IR. A non-SUSY solution with $r_e>r_s$ is shown in Figure \ref{fig:Plot1}. It is indeed very similar to the KT solution. Figure \ref{fig:Plot2} shows a non-SUSY solution with $r_e<r_s$, where $e^{2A}$ is forced to go to zero according to (\ref{eq:Exp-4Adiverges}). The pictures are plots of the solutions that asymptote to the KT background, which means that the solutions are asymptotically AdS (up to the usual logarithm).

\subsection*{Physical interpretation?}

If we ignore the issues of the singularities for a moment, we can compute physical properties of these solutions. One such a non-trivial property that we have found is that the force on a probe $p$-brane identically vanishes for all solutions that satisfy the first-order equations given by $\mathcal{S}_0$ and $\mathcal{S}_1(c)$ (or, equivalently, $\mathcal{S}_2(d)$). To see this explicitly we recall the potential on a probe $p$-brane:
\begin{equation}
\label{branepotential}
V = -T_p\int_{p+1}e^{\frac{([p-3])}{4}\phi}\sqrt{|g_{p+1}|} + Q_p\int_{p+1} C_{p+1} 
  = -T_p e^{\frac{([p-3])}{4}\phi(r)}e^{(p+1)A(r)} +Q_p\mathcal{C}(r)\, .
\end{equation}
where we introduced the function $\mathcal{C}(r)$ defined via $C_{p+1}=\mathcal{C}(r)\epsilon_{p+1}$.
In our conventions the test brane has $T_p=Q_p$. For all the solutions obeying the first-order flows derived from $\mathcal{S}_0$, $\mathcal{S}_1$ and $\mathcal{S}_2$ the following Hodge-duality relation:
\begin{equation}\label{Hodge}
\star_{9-p}H = \e^{\frac{p+1}{4}\phi}F_{6-p}
\end{equation}
holds by construction. By the virtue of the $H$ equation of motion, this implies that, up to a constant:
\begin{equation}
\mathcal{C}(r)= e^{(p+1)A + \frac{p-3}{4}\phi}\,.
\end{equation}
When this is plugged into the potential for a test D$p$ brane (\ref{branepotential}) we find that it vanishes identically. This behavior is well known for the SUSY solutions described by $\mathcal{S}_0$. What it physically means is that the charges and tension that reside in the fluxes (originating from the dissolved fractional branes) are of equal magnitude such that the gravitational attraction cancels exactly the Coulomb-like repulsion \cite{Blaback:2010sj}.  This behavior is typical for SUSY solutions and we find it surprising that it extends to these new solutions. 

Note that a second class of solutions exist in which we insert a minus sign in (\ref{Hodge}) as follows $\star_{9-p}H = -\e^{\frac{p+1}{4}\phi}F_{6-p}$. Such solutions will then possess a no-force condition on anti-Dp branes, which are defined by $T_p=-Q_p$ in our conventions. 

Concerning the singularity at small radius, we have little insight to offer. The singular behavior for $r_e>r_s$ is similar to the singular behavior of the KT solution, but, unlike the KT solution, we do not yet know whether these solutions are any physical. 
According to Gubser's criterion for ``acceptable" singularities \cite{Gubser:2000nd}, if the systems is put at finite-temperature by introducing a blackening factor into the Ansatz, the singularity should get cloaked by the horizon to be acceptable. This could be a test, which is however numerically involved, and in Section \ref{finite} we show that at least a class of finite $T$ solutions can be found from explicit first-order flows that are deformations of the first-order flows we have found here.

\subsection*{More general Ans\"atze}

Deformations of the SUSY KT background typically involve the running of a $T^{1,1}$ modulus $w$, defined as follows \cite{Buchel:2001gw}:
\begin{equation}\label{generalKT}
\d s^2_{T^{1,1}} = e^{8w}e_{\psi}^2 + e^{-2w}\Bigl( e_{\theta_{1}}^2 + e_{\phi_{1}}^2 + e_{\theta_{2}}^2 + e_{\phi_{2}}^2\Bigr)\,,
\end{equation}
with $e_{\psi}, e_{\theta_{1,2}}, e_{\phi_{1,2}}$ a standard labeling of the Maurer-Cartan forms on $T^{1,1}$. Solutions on the deformed conifold, that resolve the KT singularity, require further moduli to be turned on, as follows \cite{Klebanov:2000nc}
\begin{equation}
\d s^2_{T^{1,1}} = e^{8w}e_{\psi}^2 + e^{-2w}\Bigl( e^{y_1}[e_{\theta_{1}}^2 + e_{\phi_{1}}^2] + e^{y_2}[e_{\theta_{2}}^2 + e_{\phi_{2}}^2]\Bigr)\,.
\end{equation}
Principle functions (superpotentials) are known for these more general situations but we have not been able to find extensions that would reproduce the principle functions $\mathcal{S}_1$ or 
$\mathcal{S}_2$ of equations (\ref{S1}, \ref{S2}) upon truncation of the extra moduli.

\section{Towards first-order flows at finite temperature}\label{finite}

In this section we demonstrate how our simplified fractional brane Ans\"atze allow a finite temperature extension that is still described by a first-order flow equation. The first-order flow equations we find are restricted to the blackening of the simplified Ansatz (\ref{simpleAnsatz}) and  we have not yet been able to extend the flow to the more general Ans\"atze used in (\ref{generalKT}) for the KT black brane when $p=3$ \cite{Buchel:2001gw, Aharony:2007vg}. The latter solutions necessarily include the dynamical shape modulus $e^w$ of the $T^{1,1}$, see equation (\ref{generalKT}). Since our non-SUSY solutions have the same UV as KT we expect the singularity to only get cloaked by a finite temperature horizon if the relevant $T^{1,1}$ modulus is included.  This seems to be confirmed by a brief analysis of our first-order finite temperature flow.

\subsection{Black domain walls}

Acceptable singularities should develop a horizon at sufficiently large temperatures \cite{Gubser:2000nd}. This can be analysed by introducing a blackening factor $h^2$  in front of the time component in the metric Ansatz. As before we take the action of $D$-dimensional gravity coupled to scalar fields subject to a potential (\ref{actionscalars}) and we are interested in the following Ansatz for a domain wall at finite temperature:

\begin{equation}
ds^2_D= f(z)^2 dz^2  + g(z)^2 (-h(z)^2dt^2+d \vec{x}^2_{D-2}) \, .
\end{equation}
If we choose a field basis in which $g$ is replaced by $ G\equiv g^{D-1}h$ the effective action takes a simple form: 
\begin{equation}
S=\int \mathrm{d}z\left\{
\frac{G}{f}\left[\frac{(D-2)}{(D-1)} \left( \frac{G'^2}{G^2}-\frac{h'^2}{h^2} \right)-\frac{1}{2}h_{ab}\phi'^a\phi'^b\right]
-GfV(\phi)\right\}\,,
\label{actionFiniteT}
\end{equation}
The $h$ EOM can be integrated (in any gauge) to obtain:
\begin{equation}
\frac{Gh'}{fh}= v\,,
\end{equation}
where $v$ is constant. This shows that finite temperature extension of domain walls have a very universal character. The constant $v$ can be interpreted as a Noether charge for the radial flow as was observed earlier in \cite{Aharony:2007vg}. Here we solve for $h$ explicitly. For instance, in the gauge $f=G$ we have:
\begin{equation}
h = \exp(v z + C)\,,
\end{equation}
where $C$ a constant which we can take to be zero by shifting $z$. This eliminates one unknown completely. 
If the field $h$ is integrated out in the effective action we obtain a new effective action but this time the Hamiltonian constraint does not set the energy equal to zero but rather proportional to $v^2$, in complete analogy with the non-extremality parameter $v$ for spherical black holes.\footnote{To be more precise, the Hamiltonian constraint still sets the energy in (\ref{eq:HJ}) equal to zero, but integrating out the $h$-EOM introduces a constant term in the potential proportional to $v^2$ which can then be reinterpreted as the energy.} The analogy even extends further since, as for black holes, the non-extremality parameter turns out to be proportional to the product $s T$ with $s$ the entropy density of the domain wall when regarded as a black brane in 10D and $T$ the temperature \cite{Aharony:2007vg}.\footnote{Spherical black hole solutions of 4D gravity coupled to massless scalars and Abelian vector fields have the following metric: $\d s^2 = - \e^{2U}\d t^2 + e^{-2U}\Bigl(\e^{-4A(\tau)}\d\tau^2 + \e^{-2A(\tau)}\d\Omega_2^2\Bigr)$ where $e^A = v^{-1}\sinh(v\tau)$ and $v=2ST$. }

Let us now briefly discuss a finite temperature extensions of the non-SUSY solutions presented in section (\ref{sec:NewSol}).

\subsection{Non-SUSY at finite temperature}

Interestingly one can find explicit first-order flows for the finite $T$ solutions, similar to what has been noticed for black holes \cite{Miller:2006ay} and for black branes in flat space \cite{Janssen:2007rc}. To illustrate this we take $p=3$ and $h_{ab}$, $V(\phi)$ in (\ref{actionFiniteT}) are defined as in (\ref{action}, \ref{potential}). The Hamilton-Jacobi equation for the principle function at finite temperature is:
\begin{equation}
\frac{G^2}{3}(\partial_G\mathcal{S})^2-\frac{h^2}{3}(\partial_h\mathcal{S})^2-\frac{1}{2}(\partial_\varphi\mathcal{S})^2-\frac{1}{2}(\partial_\phi\mathcal{S})^2-\frac{1}{2}e^{-\sqrt{\frac{3}{5}}\varphi+\phi}(\partial_b\mathcal{S})^2+G^2V=0\,.
\end{equation}
We have found the following solution:
\begin{equation}
\mathcal{S}_T=G \left(	m be^{\sqrt{\frac{5}{3}}\varphi}\pm 10 e^{\frac{2}{\sqrt{15}}\varphi}\cosh \left( \frac{1}{\sqrt{5}} \left( \sqrt{3} \ln(h) \sin \alpha_1 + \sqrt{2} \phi \cos \alpha_1 \right) + \alpha_2 \right)	\right)
\label{eq:CoshBlack} \, .
\end{equation}
For $\alpha_1=0$ we recover our previous principle function at zero temperature (\ref{S1}). 
Let us now briefly discuss the finite temperature solution. We have two free parameters or constants of motion in $\mathcal{S}_T$, $\alpha_1$ and $\alpha_2$. HJ theory gives us two algebraic equations, namely the following constants of motion (\ref{eq:betas}): 
\begin{eqnarray}
\beta_1 &=& \frac{\partial \mathcal{S}_T}{\partial \alpha_1} = \pm 10G e^{\frac{2}{\sqrt{15}}\varphi} \sinh \left( \frac{1}{\sqrt{5}} \left( \sqrt{3} \ln(h) \sin \alpha_1 + \sqrt{2} \phi \cos \alpha_1 \right) + \alpha_2 \right) \cdot
\nonumber \\
 && \qquad \qquad \qquad \qquad \cdot \frac{1}{\sqrt{5}} \left( \sqrt{3} \ln(h) \cos \alpha_1 - \sqrt{2} \phi \sin \alpha_1 \right) \, ,
\\
\beta_2 &=& \frac{\partial \mathcal{S}_T}{\partial \alpha_2} = \pm  10G e^{\frac{2}{\sqrt{15}}\varphi} \sinh\left( \frac{1}{\sqrt{5}} \left( \sqrt{3} \ln(h) \sin \alpha_1 + \sqrt{2} \phi \cos \alpha_1 \right) + \alpha_2 \right) \, .
\nonumber 
\end{eqnarray}
From this we deduce that: 
\begin{equation}
h=e^{\sqrt{\frac{2}{3}} \phi \tan \alpha_1 + \frac{\beta_1}{\beta_2}} \, .
\label{SolvedBlack}
\end{equation}
If we use this relation in $\mathcal{S}_T$ then it is obvious that the remaining first-order equation are very similar to the $\mathcal{S}_1$ flow at zero temperature. The analysis of section \ref{sec:NewSol} can be repeated to find that a horizon can only form in combination with a singularity such that it cannot get cloaked from the perspective of the $\mathcal{S}_T$ flow. This is expected since the known finite temperature KT extension requires the $T^{1,1}$ modulus $e^w$ introduced in (\ref{generalKT}) to have a non-trivial flow \cite{Aharony:2007vg}.
Nevertheless, it is interesting that explicit solutions to the HJ equation can be found even at finite temperature. The non-SUSY solutions could have a finite temperature extension similar to the known KT extension if the relevant $T^{1,1}$ modulus is included. However, finding first-order equations for such systems is non-trivial and we leave this for future work.

\section{Discussion} \label{sec:concl}

We have found new SUSY-breaking solutions in type IIA/B supergravity describing throat geometries supported by fractional $p$-brane fluxes. The SUSY-breaking is such that the large radius limit of the solutions is the same as the SUSY solutions \cite{Herzog:2000rz} and  the force on a probe D$p$-brane inside the throat vanishes. Solutions of this kind could be useful as holographic backgrounds, especially when $p=3$, where the solution describes a deformation of the Klebanov-Tseytlin background \cite{Klebanov:2000nc}, dual to a cascading $SU(N)\times SU(N+M)$ SUSY gauge theory. It is of general interest to extend holographic studies of strongly coupled gauge theories to situations in which SUSY is broken. On the supergravity side this is typically done using a perturbative approach \cite{Borokhov:2002fm} that can capture small deviations from the SUSY solution. 

In this paper we have shown that one can do better and can integrate the second-order equations to  first-order integrations for deformations that preserve the no-force condition on a test D$p$-brane inside the throat. We have organised the computations using an effective one-dimensional action that is naturally obtained after reducing the fractional branes to Minkowski-sliced domain
wall solutions in $p+2$ dimensions. As a side comment we have suggested (in the Appendix) that the $AdS$-sliced domain walls of the same $(p+2)$-dimensional theories lift to a class of new AdS compactifications. Only when $p=6$ have these AdS compactifications been found explicitly \cite{Blaback:2011pn, Apruzzi:2013yva,Junghans:2014wda} and they appear to be SUSY for certain choices of integration constants \cite{Apruzzi:2013yva,Junghans:2014wda}. 

Concerning the newly found non-SUSY deformations of the fractional branes of \cite{Herzog:2000rz} we have not yet explored their physical meaning. Like the SUSY backgrounds, naked singularities arise at small distances and this could be a worry. We found two classes of solutions and for one class the singularity is in all respects the same as the KT singularity. As future work one might attempt to find a finite temperature regularisation which would make the solutions physically acceptable \cite{Gubser:2000nd}. We have initiated such a search by providing explicit solutions to the Hamilton-Jacobi equation at finite temperature. Unfortunately, our solutions most likely do not contain enough shape moduli of the $T^{1,1}$ in order to allow for smooth geometries. Another possible direction is to find a (non-supersymmetric) deformation of the $6d$ conifold geometry that renders the $10d$ solution regular exactly as it happens for the (supersymmetric) Klebanov-Strassler background. In fact, the gauge theory analysis of the non-SUSY no-force solution of \cite{Dymarsky:2011ve} shows that their supergravity background describes only one of two UV perturbations related by the $\mathcal{Z}_2$ symmetry of the dual quiver gauge theory. It is plausible that (the regular version of) the solution we found in this paper is exactly the missing gravity background mentioned in \cite{Dymarsky:2011ve}.

One can also compute correlation functions of the dual gauge theory holographically to shed light on the relevance of the new supergravity solutions we have found \cite{Argurio:2013uba, Argurio:2014rja}.

\section*{Acknowledgements}

We like to thank Riccardo Argurio and Ruben Monten for useful discussions.  BT is Aspirant FWO. TVR is supported by a Pegasus fellowship and by the Odysseus programme of the FWO. SK is supported in part by the ERC Starting Grant 240210 -- String-QCD-BH and by the John Templeton Foundation Grant 48222. We also acknowledge support from the European Science Foundation Holograv Network.

\appendix
\section{Curved domain walls as warped AdS vacua?}

In this section we consider the same $(p+2)$-dimensional theories (\ref{action}, \ref{potential}) but instead of flat domain walls we investigate AdS-sliced domain walls.

\subsection{D6 branes}

Let us first look at $6$-branes. For the ordinary $6$-brane it is known that it can be obtained as a domain wall in maximal 8-dimensional gauged SUGRA obtained from an $S^2$ reduction of the massless IIA SUGRA \cite{Salam:1984ft}. Here we discuss the extension to massive IIA which naturally leads to the fractional D6 solution which is not SUSY \cite{Janssen:1999sa} (see the explanation around equation (\ref{JMO1})). Whether the $S^2$ reduction of massive IIA also leads to a maximal gauged SUGRA is not known to us and under investigation. At this point we simply work with the bosonic Kaluza--Klein reduced theory obtained in (\ref{action}, \ref{potential}) without worrying about a possible fermionic extension that makes it into (a truncation of) a gauged supergravity in eight dimensions.    

If we then consider a domain wall Ansatz with an $AdS_7$ worldvolume then a would-be solution, defined by a specific profile for $f, g, \varphi, \phi, b$, lifts to the following $10$-dimensional solution
\begin{eqnarray}
\d s_{10}^2 &=& \e^{2\alpha\varphi(\theta)}[g^2(\theta)\d s_{AdS_7}^2 + f^2(\theta)\d\theta^2] + \e^{2\beta\varphi(\theta)}\d\Omega_2^2\,,\nonumber\\
 B_2  &=&   b(\theta)\epsilon_2\,,\\
 F_2  &=&  F_0 B_2\,, \label{AdSdomainwall}
\nonumber
\end{eqnarray}
where $\epsilon_2$ is the volume element on the $S^2$ with standard normalised metric $\d\Omega_2^2$. This Ansatz captures the $AdS_7$ compactifications discussed first in \cite{Blaback:2011nz, Blaback:2011pn}, which were further developed and properly understood in \cite{Apruzzi:2013yva} (see also \cite{Junghans:2014wda, Gaiotto:2014lca}). In the latter references the above solution was not regarded as an $AdS_7$ sliced domain wall in 8 dimensions but as a warped $AdS_7$ compactification on a compact space that is conformal to an $S^3$:
\begin{equation}
\d s_{10}^2 = \e^{2A(\theta)}\d s_{AdS_7}^2 +\e^{2B(\theta)}\Bigl[\d\theta^2 + \sin^2(\theta)\d\Omega_2^2\Bigr]\,.\label{AdScompactification}
\end{equation}
Indeed, by properly choosing the gauge for $f(\theta)$ the metric in (\ref{AdSdomainwall})  can be mapped to (\ref{AdScompactification}).

Generically the lift of an AdS-sliced domain wall in some $d$-dimensional gauged SUGRA to $10$ dimensions does not relate to a compactification because the dimensions perpendicular to the AdS slicing do not need to be compact. The fact that a compactification is possible was alluded to in \cite{Blaback:2010sj, Blaback:2011nz} where it was noticed that the presence of the Romans mass allowed the tadpole condition ($Q_6$ is the total 6-brane charge):
\begin{equation}
\int_{S^3} F_0 H = Q_{6}\,,
\end{equation}
to be satisfied. A full proof that the space was topologically compact was presented in \cite{Apruzzi:2013yva} by finding first-order equations for the solution instead of second-order equations allowing a numerical treatment of the solutions. The first-order integration was possible since \cite{Apruzzi:2013yva} demonstrated, contrary to the claim in \cite{Blaback:2011nz}, that such an Ansatz could lead to SUSY solutions.\footnote{
Apart from the SUSY solutions, reference \cite{Junghans:2014wda} has given evidence for a 1-parameter family of non-SUSY solutions. 
}
If the $8$-dimensional theory can be seen as a consistent truncation of an $8$-dimensional maximal SUGRA then we furthermore conjecture that the domain wall is also 1/2 SUSY from the point of view of the $8$-dimensional theory.  Hence, from the point of view of the curved domain wall, one expects to have a superpotential that provides the first-order equations for \emph{curved} domain walls, see for instance \cite{Freedman:2003ax}. 

So far we have hinted that the $AdS_7$ solutions constructed entirely from a $10$-dimensional viewpoint can be possibly described within $8$-dimensional gauged supergravity as a (SUSY?) AdS-sliced domain wall solution. Nonetheless the natural lower-dimensional supergravity to look at should be $7$-dimensional half-maximal gauged SUGRA in which the $AdS_7$ is the natural vacuum.
It became a standard lore that flux compactifications to (SUSY) AdS vacua with space-filling branes lead to gauged supergravity theories in the lower dimension, \emph{if} the branes are calibrated and smeared over the internal space \cite{Blaback:2010sj}. The smearing procedure implies constant dilaton and the absence of warping. After such a smearing the AdS solution simplifies to $AdS_7 \times S^3$ \cite{Blaback:2010sj} since both $e^A$ and $e^B$ become constant in that limit and $F_2$ vanishes. Since $S^3$ is a group manifold one furthermore expects that the smeared compactification gives rise to a half-maximal SUGRA in $7$ dimensions.  Reference \cite{Danielsson:2013qfa} has demonstrated that this belief is not founded since this  $AdS_7$ solution is a counterexample.\footnote{However, reference \cite{Karndumri:2014hma} mentions that the 7-dimensional gauged SUGRA might still exist.}

\subsection{D\emph{p} branes}

As mentioned above, when the $6$-branes are smeared the solution becomes a simple unwarped $AdS_7\times S^3$ solution. Similar vacua can be made in any dimension less than $7$ as well \cite{Silverstein:2004id,Blaback:2010sj}: when a smeared D$p$-brane charge is cancelled by a combination of $H$ flux filling some $S^3$ and $F_{6-p}$ flux filling an orthogonal $S^{6-p}$ then one can solve the $10d$ equations of motion if the lower-dimensional part is $AdS_{p+1}$.\footnote{Only for $p=5$ does this fails at first sight since the absence of curvature of the $S^1$ factor makes the Einstein equation in that direction inconsistent. However, one could expect that a certain fibration of $S^1$ over $S^3$ could work.} The fully localised (i.e. back-reacted) solutions are probably out of reach since the equations for the warp factor, conformal metric factors, dilaton, and $B$-field profile will not be ODEs any more when $p\neq 6$. However, once the D$p$ branes are smeared over the $S^{6-p}$ one ends up in a situation which is practically analogues to the D$6$ $AdS_7$ solution discussed above. Such a set-up should be described by the following Ansatz:
\begin{eqnarray}
\d s_{10}^2 &=& \e^{2A(\theta)}\d s_{AdS_{p+1}}^2 +\e^{2B(\theta)}\d\Omega_{6-p}^2 + \e^{2C(\theta)}\Bigl[\d\theta^2 + \sin^2(\theta)\d\Omega_2^2\Bigr]\,,\nonumber\\
F_{8-p} &=& B_2\wedge F_{6-p} \,,\nonumber\\
B_2 &=& b(\theta)\epsilon_2\,,\\
F_{6-p} &=& m \epsilon_{6-p}\,,\nonumber
\end{eqnarray}  
where $A, B, C, b, \phi$ all only depend on $\theta$. The solutions are currently under investigation. Clearly also here the solutions can be described by an AdS-sliced domain wall after compactification over the $S^{6-p}$ and further over the $S^2$ inside the $S^3$. 

For these $AdS_{p+1}$ solutions, we anticipate physics which is similar to the $AdS_7$ case \cite{Blaback:2011nz, Apruzzi:2013yva, Gaiotto:2014lca, Junghans:2014wda}. The arguments of \cite{Blaback:2011nz} can directly be used in this situation to demonstrate that the solution should have a divergent $e^{-\phi}H^2$ density. From the analogy with the $AdS_7$ solution we expect that the singularity gets resolved by polarising the $p$-brane into a spherical $(p+2)$-brane wrapping a finite size 2-sphere inside the $S^3$. The non-compact ``anti-brane" solutions, that have a flat worldvolume Mink$_{p+1}$ instead of $AdS_{p+1}$ cannot be stabilised at finite radius and hence the singularity cannot be resolved that way \cite{Bena:2012tx, Bena:2012vz, Bena:2014bxa,Bena:2014jaa}.

The case $p=3$ is somewhat more intriguing. There one should be able to S-dualise the solution and smear the 3-brane over the $S^3$ filled with $F_3$ flux instead of $H_3$ flux.  The $3$-form singularity in this case might get resolved by a spherical NS$5$ brane. This leads us to further speculate that the solution with fully-localised $3$-branes, which might be out of reach, would be similar to the (also hypothetical) Polchinski--Strassler (PS) solution \cite{Polchinski:2000uf}  where a web of spherical D5/NS5 branes wrap 2 cycles inside the $S^5$ of the deformed $AdS_5 \times S^5$. The crucial difference with the PS configuration is that the (deformed) $AdS_5$ part of the PS solution is replaced by an $AdS_4$-sliced domain wall and the $S^5$ is replaced by some fibration of $S^2$ over $S^3$. The space transversal to the $AdS_4$ is also compact. In that sense one could think of the PS solution in terms of a flat domain wall in $5$ dimensions where the space transversal to the wall is non-compact whereas the new solution is an AdS-sliced domain wall in $5$ dimensions with a compact space transversal to the $AdS_4$ wall.

Similar results are expected for M2 branes with $AdS_3$ worldvolume \cite{Blaback:2013hqa}, that can be regarded as a compactification of 11-dimensional supergravity down to 3 dimensions.

\bibliography{refs}

\providecommand{\href}[2]{#2}\begingroup\raggedright\begin{thebibliography}{10}

\bibitem{Maldacena:2001pb}
J.~M. Maldacena and H.~S. Nastase,  {\em {The Supergravity dual of a theory
  with dynamical supersymmetry breaking}}, JHEP {\bf 0109} (2001) 024
[\href{http://www.arXiv.org/abs/hep-th/0105049}{{\tt hep-th/0105049}}].

\bibitem{Kachru:2002gs}
S.~Kachru, J.~Pearson and H.~L. Verlinde,  {\em {Brane / flux annihilation and
  the string dual of a nonsupersymmetric field theory}}, JHEP {\bf 0206} (2002)
  021
[\href{http://www.arXiv.org/abs/hep-th/0112197}{{\tt hep-th/0112197}}].

\bibitem{Argurio:2006ny}
R.~Argurio, M.~Bertolini, S.~Franco and S.~Kachru,  {\em {Gauge/gravity duality
  and meta-stable dynamical supersymmetry breaking}}, JHEP {\bf 0701} (2007)
  083
[\href{http://www.arXiv.org/abs/hep-th/0610212}{{\tt hep-th/0610212}}].

\bibitem{Kachru:2003aw}
S.~Kachru, R.~Kallosh, A.~D. Linde and S.~P. Trivedi,  {\em {De Sitter vacua in
  string theory}}, Phys.Rev. {\bf D68} (2003) 046005
[\href{http://www.arXiv.org/abs/hep-th/0301240}{{\tt hep-th/0301240}}].

\bibitem{Bena:2009xk}
I.~Bena, M.~Grana and N.~Halmagyi,  {\em {On the Existence of Meta-stable Vacua
  in Klebanov-Strassler}}, JHEP {\bf 1009} (2010) 087
[\href{http://www.arXiv.org/abs/0912.3519}{{\tt 0912.3519}}].

\bibitem{McGuirk:2009xx}
P.~McGuirk, G.~Shiu and Y.~Sumitomo,  {\em {Non-supersymmetric infrared
  perturbations to the warped deformed conifold}}, Nucl.Phys. {\bf B842} (2011)
  383--413
[\href{http://www.arXiv.org/abs/0910.4581}{{\tt 0910.4581}}].

\bibitem{Bena:2012bk}
I.~Bena, M.~Grana, S.~Kuperstein and S.~Massai,  {\em {Anti-D3 Branes: Singular
  to the bitter end}}, Phys.Rev. {\bf D87} (2013), no.~10, 106010
[\href{http://www.arXiv.org/abs/1206.6369}{{\tt 1206.6369}}].

\bibitem{Blaback:2010sj}
J.~Blaback, U.~H. Danielsson, D.~Junghans, T.~Van~Riet, T.~Wrase and
  M.~Zagermann,  {\em {Smeared versus localised sources in flux
  compactifications}}, JHEP {\bf 1012} (2010) 043
[\href{http://www.arXiv.org/abs/1009.1877}{{\tt 1009.1877}}].

\bibitem{Blaback:2011nz}
J.~Blaback, U.~H. Danielsson, D.~Junghans, T.~Van~Riet, T.~Wrase and
  M.~Zagermann,  {\em {The problematic backreaction of SUSY-breaking branes}},
  JHEP {\bf 1108} (2011) 105
[\href{http://www.arXiv.org/abs/1105.4879}{{\tt 1105.4879}}].

\bibitem{DeWolfe:2004qx}
O.~DeWolfe, S.~Kachru and H.~L. Verlinde,  {\em {The Giant inflaton}}, JHEP
  {\bf 0405} (2004) 017
[\href{http://www.arXiv.org/abs/hep-th/0403123}{{\tt hep-th/0403123}}].

\bibitem{Blaback:2012nf}
J.~Blaback, U.~H. Danielsson and T.~Van~Riet,  {\em {Resolving anti-brane
  singularities through time-dependence}}, JHEP {\bf 1302} (2013) 061
[\href{http://www.arXiv.org/abs/1202.1132}{{\tt 1202.1132}}].

\bibitem{Danielsson:2014yga}
U.~H. Danielsson and T.~Van~Riet,  {\em {Fatal attraction: more on decaying
  anti-branes}},
\href{http://www.arXiv.org/abs/1410.8476}{{\tt 1410.8476}}.

\bibitem{Klebanov:2000nc}
I.~R. Klebanov and A.~A. Tseytlin,  {\em {Gravity duals of supersymmetric SU(N)
  x SU(N+M) gauge theories}}, Nucl.Phys. {\bf B578} (2000) 123--138
[\href{http://www.arXiv.org/abs/hep-th/0002159}{{\tt hep-th/0002159}}].

\bibitem{Klebanov:2000hb}
I.~R. Klebanov and M.~J. Strassler,  {\em {Supergravity and a confining gauge
  theory: Duality cascades and chi SB resolution of naked singularities}}, JHEP
  {\bf 0008} (2000) 052
[\href{http://www.arXiv.org/abs/hep-th/0007191}{{\tt hep-th/0007191}}].

\bibitem{Aharony:2007vg}
O.~Aharony, A.~Buchel and P.~Kerner,  {\em {The Black hole in the throat:
  Thermodynamics of strongly coupled cascading gauge theories}}, Phys.Rev. {\bf
  D76} (2007) 086005
[\href{http://www.arXiv.org/abs/0706.1768}{{\tt 0706.1768}}].

\bibitem{Dymarsky:2011ve}
A.~Dymarsky and S.~Kuperstein,  {\em {Non-supersymmetric Conifold}}, JHEP {\bf
  1208} (2012) 033
[\href{http://www.arXiv.org/abs/1111.1731}{{\tt 1111.1731}}].

\bibitem{deBoer:1999xf}
J.~de~Boer, E.~P. Verlinde and H.~L. Verlinde,  {\em {On the holographic
  renormalization group}}, JHEP {\bf 0008} (2000) 003
[\href{http://www.arXiv.org/abs/hep-th/9912012}{{\tt hep-th/9912012}}].

\bibitem{Skenderis:2006rr}
K.~Skenderis and P.~K. Townsend,  {\em {Hamilton-Jacobi method for curved
  domain walls and cosmologies}}, Phys.Rev. {\bf D74} (2006) 125008
[\href{http://www.arXiv.org/abs/hep-th/0609056}{{\tt hep-th/0609056}}].

\bibitem{Andrianopoli:2009je}
L.~Andrianopoli, R.~D'Auria, E.~Orazi and M.~Trigiante,  {\em {First Order
  Description of D=4 static Black Holes and the Hamilton-Jacobi equation}},
  Nucl.Phys. {\bf B833} (2010) 1--16
[\href{http://www.arXiv.org/abs/0905.3938}{{\tt 0905.3938}}].

\bibitem{Trigiante:2012eb}
M.~Trigiante, T.~Van~Riet and B.~Vercnocke,  {\em {Fake supersymmetry versus
  Hamilton-Jacobi}}, JHEP {\bf 1205} (2012) 078
[\href{http://www.arXiv.org/abs/1203.3194}{{\tt 1203.3194}}].

\bibitem{Janssen:2007rc}
B.~Janssen, P.~Smyth, T.~Van~Riet and B.~Vercnocke,  {\em {A First-order
  formalism for timelike and spacelike brane solutions}}, JHEP {\bf 0804}
  (2008) 007
[\href{http://www.arXiv.org/abs/0712.2808}{{\tt 0712.2808}}].

\bibitem{DeWolfe:1999cp}
O.~DeWolfe, D.~Freedman, S.~Gubser and A.~Karch,  {\em {Modeling the
  fifth-dimension with scalars and gravity}}, Phys.Rev. {\bf D62} (2000) 046008
[\href{http://www.arXiv.org/abs/hep-th/9909134}{{\tt hep-th/9909134}}].

\bibitem{future}
B.~Truijen and T.~Van~Riet,  {\em {What is fake supersymmetry really?}}, to
  appear.

\bibitem{Blaback:2012mu}
J.~Blaback, B.~Janssen, T.~Van~Riet and B.~Vercnocke,  {\em {Fractional branes,
  warped compactifications and backreacted orientifold planes}}, JHEP {\bf
  1210} (2012) 139
[\href{http://www.arXiv.org/abs/1207.0814}{{\tt 1207.0814}}].

\bibitem{Herzog:2000rz}
C.~P. Herzog and I.~R. Klebanov,  {\em {Gravity duals of fractional branes in
  various dimensions}}, Phys.Rev. {\bf D63} (2001) 126005
[\href{http://www.arXiv.org/abs/hep-th/0101020}{{\tt hep-th/0101020}}].

\bibitem{Janssen:1999sa}
B.~Janssen, P.~Meessen and T.~Ortin,  {\em {The D8-brane tied up: String and
  brane solutions in massive type IIA supergravity}}, Phys.Lett. {\bf B453}
  (1999) 229--236
[\href{http://www.arXiv.org/abs/hep-th/9901078}{{\tt hep-th/9901078}}].

\bibitem{Freedman:2003ax}
D.~Freedman, C.~Nunez, M.~Schnabl and K.~Skenderis,  {\em {Fake supergravity
  and domain wall stability}}, Phys.Rev. {\bf D69} (2004) 104027
[\href{http://www.arXiv.org/abs/hep-th/0312055}{{\tt hep-th/0312055}}].

\bibitem{Gubser:2000nd}
S.~S. Gubser,  {\em {Curvature singularities: The Good, the bad, and the
  naked}}, Adv.Theor.Math.Phys. {\bf 4} (2000) 679--745
[\href{http://www.arXiv.org/abs/hep-th/0002160}{{\tt hep-th/0002160}}].

\bibitem{Bergshoeff:2004nq}
E.~Bergshoeff, M.~Nielsen and D.~Roest,  {\em {The Domain walls of gauged
  maximal supergravities and their M-theory origin}}, JHEP {\bf 0407} (2004)
  006
[\href{http://www.arXiv.org/abs/hep-th/0404100}{{\tt hep-th/0404100}}].

\bibitem{Apruzzi:2013yva}
F.~Apruzzi, M.~Fazzi, D.~Rosa and A.~Tomasiello,  {\em {All $AdS_7$ solutions
  of type II supergravity}}, JHEP {\bf 1404} (2014) 064
[\href{http://www.arXiv.org/abs/1309.2949}{{\tt 1309.2949}}].

\bibitem{Blaback:2011pn}
J.~Blaback, U.~H. Danielsson, D.~Junghans, T.~Van~Riet, T.~Wrase and
  M.~Zagermann,  {\em {(Anti-)Brane backreaction beyond perturbation theory}},
  JHEP {\bf 1202} (2012) 025
[\href{http://www.arXiv.org/abs/1111.2605}{{\tt 1111.2605}}].

\bibitem{Junghans:2014wda}
D.~Junghans, D.~Schmidt and M.~Zagermann,  {\em {Curvature-induced Resolution
  of Anti-brane Singularities}},
\href{http://www.arXiv.org/abs/1402.6040}{{\tt 1402.6040}}.

\bibitem{Gaiotto:2014lca}
D.~Gaiotto and A.~Tomasiello,  {\em {Holography for (1,0) theories in six
  dimensions}},
\href{http://www.arXiv.org/abs/1404.0711}{{\tt 1404.0711}}.

\bibitem{Danielsson:2013qfa}
U.~Danielsson, G.~Dibitetto, M.~Fazzi and T.~Van~Riet,  {\em {A note on smeared
  branes in flux vacua and gauged supergravity}}, JHEP {\bf 1404} (2014) 025
[\href{http://www.arXiv.org/abs/1311.6470}{{\tt 1311.6470}}].

\bibitem{Lu:1996hh}
H.~Lu, C.~Pope and K.~Xu,  {\em {Liouville and Toda solutions of M theory}},
  Mod.Phys.Lett. {\bf A11} (1996) 1785--1796
[\href{http://www.arXiv.org/abs/hep-th/9604058}{{\tt hep-th/9604058}}].

\bibitem{Gubser:1999pk}
S.~S. Gubser,  {\em {Dilaton driven confinement}},
\href{http://www.arXiv.org/abs/hep-th/9902155}{{\tt hep-th/9902155}}.

\bibitem{Buchel:2001gw}
A.~Buchel, C.~Herzog, I.~R. Klebanov, L.~A. Pando~Zayas and A.~A. Tseytlin,
  {\em {Nonextremal gravity duals for fractional D-3 branes on the conifold}},
  JHEP {\bf 0104} (2001) 033
[\href{http://www.arXiv.org/abs/hep-th/0102105}{{\tt hep-th/0102105}}].

\bibitem{Miller:2006ay}
C.~M. Miller, K.~Schalm and E.~J. Weinberg,  {\em {Nonextremal black holes are
  BPS}}, Phys.Rev. {\bf D76} (2007) 044001
[\href{http://www.arXiv.org/abs/hep-th/0612308}{{\tt hep-th/0612308}}].

\bibitem{Borokhov:2002fm}
V.~Borokhov and S.~S. Gubser,  {\em {Nonsupersymmetric deformations of the dual
  of a confining gauge theory}}, JHEP {\bf 0305} (2003) 034
[\href{http://www.arXiv.org/abs/hep-th/0206098}{{\tt hep-th/0206098}}].

\bibitem{Argurio:2013uba}
R.~Argurio, M.~Bertolini, L.~Pietro, F.~Porri and D.~Redigolo,  {\em
  {Supercurrent multiplet correlators at weak and strong coupling}}, JHEP {\bf
  1404} (2014) 123
[\href{http://www.arXiv.org/abs/1310.6897}{{\tt 1310.6897}}].

\bibitem{Argurio:2014rja}
R.~Argurio, D.~Musso and D.~Redigolo,  {\em {Anatomy of new SUSY breaking
  holographic RG flows}},
\href{http://www.arXiv.org/abs/1411.2658}{{\tt 1411.2658}}.

\bibitem{Salam:1984ft}
A.~Salam and E.~Sezgin,  {\em {d = 8 supergravity}}, Nucl.Phys. {\bf B258}
  (1985)
284.

\bibitem{Karndumri:2014hma}
P.~Karndumri,  {\em {RG flows in 6D N=(1,0) SCFT from SO(4) half-maximal 7D
  gauged supergravity}},
\href{http://www.arXiv.org/abs/1404.0183}{{\tt 1404.0183}}.

\bibitem{Silverstein:2004id}
E.~Silverstein,  {\em {TASI / PiTP / ISS lectures on moduli and microphysics}},
\href{http://www.arXiv.org/abs/hep-th/0405068}{{\tt hep-th/0405068}}.

\bibitem{Bena:2012tx}
I.~Bena, D.~Junghans, S.~Kuperstein, T.~Van~Riet, T.~Wrase and M.~Zagermann,
  {\em {Persistent anti-brane singularities}}, JHEP {\bf 1210} (2012) 078
[\href{http://www.arXiv.org/abs/1205.1798}{{\tt 1205.1798}}].

\bibitem{Bena:2012vz}
I.~Bena, M.~Grana, S.~Kuperstein and S.~Massai,  {\em {Polchinski-Strassler
  does not uplift Klebanov-Strassler}}, JHEP {\bf 1309} (2013) 142
[\href{http://www.arXiv.org/abs/1212.4828}{{\tt 1212.4828}}].

\bibitem{Bena:2014bxa}
I.~Bena, M.~Grana, S.~Kuperstein and S.~Massai,  {\em {Tachyonic Anti-M2
  Branes}},
\href{http://www.arXiv.org/abs/1402.2294}{{\tt 1402.2294}}.

\bibitem{Bena:2014jaa}
I.~Bena, M.~Grana, S.~Kuperstein and S.~Massai,  {\em {Giant Tachyons in the
  Landscape}},
\href{http://www.arXiv.org/abs/1410.7776}{{\tt 1410.7776}}.

\bibitem{Polchinski:2000uf}
J.~Polchinski and M.~J. Strassler,  {\em {The String dual of a confining
  four-dimensional gauge theory}},
\href{http://www.arXiv.org/abs/hep-th/0003136}{{\tt hep-th/0003136}}.

\bibitem{Blaback:2013hqa}
J.~Blaback,  {\em {A note on M2-branes in opposite charge}},
\href{http://www.arXiv.org/abs/1309.2640}{{\tt 1309.2640}}.

\end{thebibliography}\endgroup

\bibliographystyle{utphysmodb}
\end{document}